# Layer-Resolved Ultrafast XUV Measurement of Hole Transport in a Ni-TiO$_2$-Si Photoanode

Scott K. Cushing[1], Ilana J. Porter[2,3], Bethany R. Lamoureux[2], Angela Lee[2], Brett M. Marsh[2], Szilard Szoke[1], Mihai E. Vaida[2,5], Stephen R. Leone[2,3,4*]

[1]Division of Chemistry and Chemical Engineering, California Institute of Technology, Pasadena, CA 91125, USA

[2]Department of Chemistry, University of California, Berkeley, CA 94720, USA.

[3]Chemical Sciences Division, Lawrence Berkeley National Laboratory, Berkeley, CA 94720, USA.

[4]Department of Physics, University of California, Berkeley, CA 94720, USA.

[5]Department of Physics, University of Central Florida, FL 32816, USA.

Author e-mail address: srl@berkeley.edu

**Metal-oxide-semiconductor junctions are central to most electronic and optoelectronic devices. Here, the element-specificity of broadband extreme ultraviolet (XUV) ultrafast pulses is used to measure the charge transport and recombination kinetics in each layer of a Ni-TiO$_2$-Si junction. After photoexcitation of silicon, holes are inferred to transport from Si to Ni ballistically in ~100 fs, resulting in spectral shifts in the Ni M$_{2,3}$ XUV edge that are characteristic of holes and the absence of holes initially in TiO$_2$. Meanwhile, the electrons are observed to remain on Si. After picoseconds, the transient hole population on Ni is observed to back-diffuse through the TiO$_2$, shifting the Ti spectrum to higher oxidation state, followed by electron-hole recombination at the Si-TiO$_2$ interface and in the Si bulk. Electrical properties, such as the hole diffusion constant in TiO$_2$ and the initial hole mobility in Si, are fit from these transient spectra and match well with values reported previously.**

## Introduction

Metal-oxide-semiconductor (MOS) junctions are foundational to electronic devices. In solar energy research, oxide-passivated junctions have led to record photoconversion efficiencies for semiconductor solar cells and photoelectrodes[1–3]. In a solar cell, one role of the metal-oxide-passivated junction is to control surface recombination velocities, slowing the recombination at the metal-semiconductor contacts[4]. In photoelectrochemical cells, the metal oxide layer also acts as a corrosion barrier[5–9]. Even in solar photocatalytic reduction of $CO_2$, the MOS junction has proven critical by acting as a proton transport layer[10,11]. Despite the critical applications of MOS junctions, the femtosecond to picosecond charge transfer processes that occur within a photo-initiated MOS junction are still debated. It is established that field-induced tunneling dominates thin <5 nm junctions. However, in photoelectrochemical junctions, thicker barriers often lead to better performance[5]. Additionally, an amorphous or defect-rich oxide will often outperform a crystalline material. It has therefore been proposed that in p-type MOS junctions with $TiO_2$, the $Ti^{3+}$ defect states support efficient hole or proton conduction[6,12,13].

Ultrafast x-ray studies have recently brought element-specificity to time-resolved dynamics[14,15]. One approach to producing ultrafast x-ray probes is using high-harmonic generation. In high-harmonic generation, extreme ultraviolet (XUV) or soft x-ray pulses are produced by a noble gas using a table-top laser[16]. The broadband x-ray pulses can have a bandwidth of 10-100 eV, allowing for multiple elements to be temporally correlated. However, interpreting the measured x-ray dynamics in terms of ground state electronic properties is made difficult by the positive core-hole that is produced by the core-level probe transition. Advances in theory have led to accurate interpretation of atomic and molecular dynamics, but the many-body state created by the core-hole in a solid makes the theoretical interpretation of the spectral features challenging[17,18]. Nevertheless, recent progress in approximate methods has led to extraction of the electron and hole dynamics in semiconductors. The fit dynamics accurately correspond to scattering pathways within the material's band structure[19–22].

Here, we use the element-specificity of transient XUV spectroscopy to measure the charge transfer kinetics of a Ni-$TiO_2$-Si junction with band alignment as shown in Figure 1A. First, a near-infrared pump photoexcites the 200 nm thick Si in the MOS junction. Then, a broadband XUV pulse created by high harmonic generation in Ar (30 eV – 60 eV) or He (60 - 110 eV) probes the photoexcited changes in the Ti $M_{2,3}$ edge at 33 eV, the Ni $M_{2,3}$ edge at 66 and 68 eV, and the Si $L_{2,3}$ edge at 100 eV from femtoseconds to 200 ps. The increased absorption feature normally ascribed to photoexcited holes in pure Si is not measured immediately in the junction. The electron spectral feature of Si is consistent with photoexcited Si alone. A subsequent negative shift in the Ni edge energy is measured within 100 fs, which is attributed to excess holes opening up transitions to states below the Fermi energy. No change is measured for the Ti edge on this short time scale. On a picoseconds time scale, the Ni edge energy returns to its equilibrium value while a positive energy shift of the Ti edge is measured, attributed to an increase in the Ti oxidation state from back-diffusion of excess holes. On a tens to hundreds of picoseconds timescale, the Ti peak shift reaches its maximum, and then it begins to decay in-sync with the decay of the electron signature in Si.

The initial hole transport from Si to Ni in the junction is fit and compared with the fitted hole transport time in a Ni-Si sample. The average hole transit time is delayed in the junction by $33\pm2$ fs, and the hole quantum yield is measured to be $42\pm6\%$. Accounting for the $TiO_2$ thickness, the hole tunneling velocity in the $TiO_2$ is $5.8\pm0.4x10^7$ cm/s. For the $1.5x10^5$ V/cm built-in electric field of the MOS junction, this implies a $380\pm40$ cm$^2$/V•s hole mobility in Si, which matches measured Si hole mobilities. The value suggests ballistic transfer because it is unchanged by scattering in the $TiO_2$. After the tunneling process, the holes back-transfer through the $TiO_2$ with a fit diffusion constant of $1.2\pm0.1$ cm$^2$/s. The recombination of holes in the $TiO_2$ with non-transferred electrons in the Si, or injection of holes from $TiO_2$ to Si, is fit to a surface recombination/injection velocity of >200 cm/s.

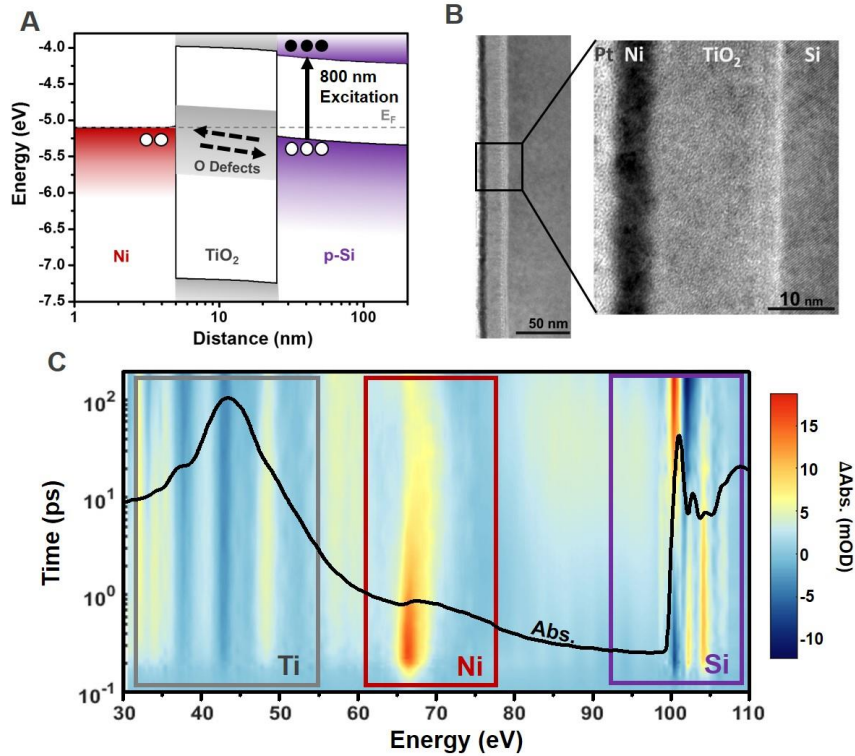

**Figure 1**. **Characterization and measurement of the Ni-TiO₂-Si junction. A)** The energy level alignment for the metal, oxide, and semiconductor is shown, along with the expected photoexcited hole transfer in the p-type MOS junction. The Si is p-type doped by boron at $10^{15}/cm^3$. The presence of oxygen defect levels (n-type) in the $TiO_2$ layer was previously confirmed by photoemission spectroscopy of a $Si-TiO_2$ junction[23]. The band bending is calculated using the drift-diffusion equation[24]. **B)** A TEM measurement of the thickness of the $TiO_2$ and Ni, which are 19±0.6 nm and 5.6±0.6 nm, respectively. The $TiO_2$ is amorphous, and an ~1 nm $SiO_2$ interface is measured where the $TiO_2$ and Si contact. **C)** The black line overlay is the ground-state XUV absorption. The regions that correspond to the Ti $M_{2,3}$ edge, the Ni $M_{2,3}$ edge, and the Si $L_{2,3}$ edge are indicated by the colored boxes. The differential XUV absorption that results from photoexcitation is shown as the background color map, with the scale on the right of the graph. Each peak is observed to have a unique response to the photo-initiated charge transfer, which will be discussed in the main text.

## Results

The Ni-TiO₂-Si thin films (Figure 1A) were prepared via physical vapor deposition under ultrahigh vacuum conditions. The 19±0.6 nm of TiO₂ and 5.6±0.6 nm of Ni were grown on a 200 nm thick p-doped (B at $10^{15}/cm^3$) Si membrane. The surface oxide of the Si could not be etched due to the fragility of the membrane. The ~1 nm SiO₂ barrier can be seen in the transmission electron microscope (TEM) image of Figure 1B and may act as a thin tunneling barrier; its presence is consistent with previous Ni-TiO$_2$-Si photoanode studies[5,25]. The TEM cross-section confirms the amorphous nature of the TiO$_2$. For this sample, the presence of the oxygen defect levels in the TiO$_2$ was previously quantified during growth of the Si-TiO$_2$ junction[23]. The resulting Ti$^{3+}$ defect states in the amorphous TiO$_2$ are generally accepted to be >1 eV below the conduction band of TiO$_2$ and span a 1-2 eV range[13,26,27]. Note that in a Si-TiO$_2$ junction alone, the band energetics promote electron transport[28], whereas in the Ni-TiO$_2$-Si junction, the band energetics create an internal field that promotes hole transport, as shown in Figure 1A. The topmost Pt layer was used for TEM imaging purposes. The band bending in Figure 1A is calculated based on a drift-diffusion model using the experimentally measured thicknesses[24].

The ground state XUV absorption (black line overlay in Figure 1C) contains the Ti M$_{2,3}$ edge at 32.6 eV, the Ni M$_{2,3}$ edges at 66.2 and 68 eV, and the Si L$_{2,3}$ edges at 99.2 and 99.8 eV. The 30-150 eV XUV range is created by high-harmonic generation in Ar or He. An Al or Zr metal filter prevents second-order diffraction modes from being observed at the XUV camera. The residual 800 nm driving laser for the HHG process is removed using a microchannel plate filter before the camera[29]. The full details of the technique are found in the SI. The ground state absorption amplitude for each elemental edge is determined by the film thickness (Figure 1B) and the absorption transition probability of the element. This is why the absorption magnitude of the ~20 nm TiO$_2$ film, which has 10 empty 3d levels, is larger than the 200 nm thick Si, which has partially occupied and s-p hybridized valence bands. Similarly, the Ni M$_{2,3}$ edge has the smallest amplitude in the static spectra because only two unoccupied 3d orbitals can absorb the XUV radiation in the relatively thin ~5 nm film.

Charge transfer in the MOS junction is photo-initiated by a 50 fs 800 nm laser pulse from a Ti:Sapphire regenerative amplifier. The Si thin film primarily absorbs the 800 nm radiation, photoexciting electrons and holes by an indirect transition to the $\Delta$ valley. The photoexcitation density is restricted to $1\times10^{20}$ carriers/cm$^3$ to minimize multiphoton absorption by using 800 nm pulse energy densities of approximately 2 mJ cm$^{-2}$. The resulting changes to the XUV absorption spectrum are shown as the background color map of Figure 1C. The blue color represents a decrease in absorption after photoexcitation, while the red color indicates an increase in absorption. The change in absorption is displayed from tens of femtoseconds to 200 picoseconds on a logarithmic time scale. The time scale is offset for visualization, with zero delay between pump and probe occurring at 100 fs. The absorption features in Figure 1C and their evolution in time represent the underlying photoexcitation, charge transfer, and heat transfer processes in the junction.

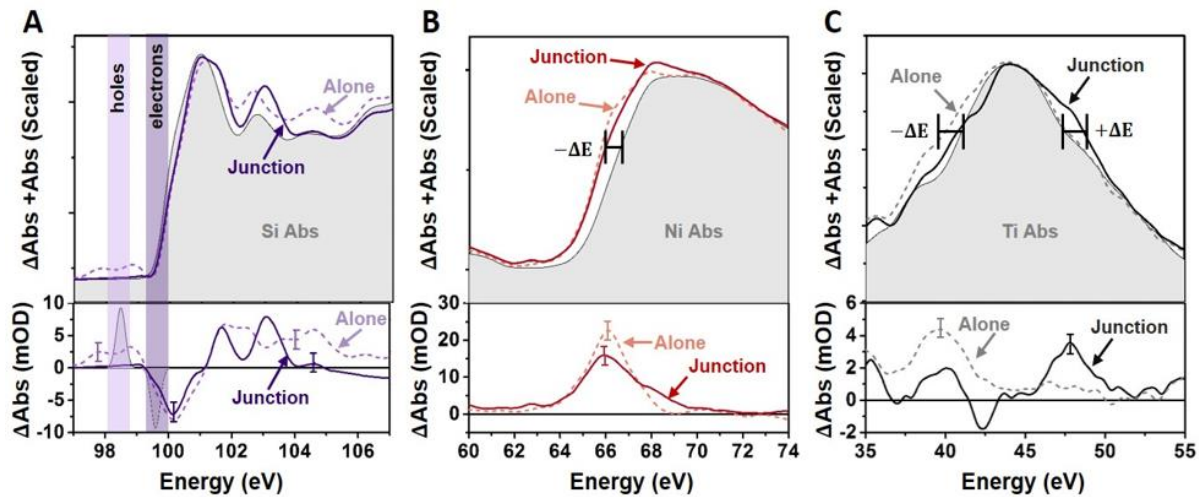

**Figure 2. Photoexcited Changes in the Si, TiO₂, and Ni separately and in a junction.** The differential absorption versus photon energy from Figure 1C are plotted in the bottom row while the top row shows the differential absorption scaled and added to the ground state absorption. The solid lines represent the excited state change for the elements in the junction for **A**) Si 100 fs after photoexcitation, **B**) Ni 100 fs after photoexcitation of the Si, and **C**) TiO₂ 1 ps after the photoexcitation of Si. The same time differential absorption versus photon energy is shown for photoexcitation of each element alone as a dashed line. As discussed in the text, different excitation wavelengths and thicknesses were required to photoexcite the elements alone as compared to the junction. A representative error bar of the experimental 2 mOD signal to noise ratio is shown at key comparison energies.

The differential absorption versus photon energy of Si and Ni at 100 fs and Ti at 1 ps is shown in the bottom row of Figure 2. The top row of Figure 2 shows the ground state absorption plus the differential absorption, scaled for visualization. This information is plotted for each element in the junction (solid lines) as well as for Si, Ni, and $TiO_2$ photoexcited on their own (dashed lines). The full differential absorption plots for the elements alone are shown in Supplementary Figure 1. For the Ni and $TiO_2$ alone, conditions are arranged to obtain similar transient absorbances as for the junction. To measure $TiO_2$ alone, a $TiO_2$ layer on a diamond substrate is measured and the pump wavelength is changed to 266 nm (approximately 1 mJ cm$^{-2}$) to excite above the ~3.2 eV band gap[30]. The 266 nm single photon absorption probability of $TiO_2$ is $10^5$ times larger than the three photon absorption at 800 nm needed to excite above the band gap, so any excitation of $TiO_2$ by 800 nm light can be neglected in the junction. To measure Ni alone, a diamond substrate supporting a thicker layer of Ni with an absorption of 0.6 OD from the pre-edge to peak is used (Supplementary Figure 2). The Ni in the junction has a 0.03 OD absorption magnitude from the pre-edge to the peak. When the thicker Ni is photoexcited with 800 nm light (approximately 0.3 mJ cm$^{-2}$), a change in absorption of ~25 mOD is measured. When the thinner Ni is excited by the same density of 800 nm light, no signal is observed within the <5 mOD experimental noise (see Supplementary Figure 2), as would be expected from the 20x decrease in absorbed 800 nm light. Additionally, an immediate rise time is not measured for the Ni in the junction as it is for Ni alone as discussed in Figure 3b.

The core-hole excited by the XUV transition perturbs the final state in the core-level transition, masking the ground state density of states. The strength of the core-hole interaction depends on the element's orbital occupation and bonding. The stronger the core-hole interaction, the more the ground state density of states is masked, and thus the information that can be obtained from the measured photoexcited state is altered. For example, the core-hole in Si is well-screened, and the critical points are only slightly shifted from the ground state band structure (Supplementary Figure 3A). Changes in the $L_{2,3}$ edge are therefore representative of the underlying carrier and lattice dynamics, and not simply due to changes in the core-hole perturbation, as shown by the $L_{2,3}$ edge absorption of different Si oxidation states in Supplementary Figure 4A. After photoexcitation, the differential absorption features above 101 eV are known to correspond to structural changes[31]. Below 101 eV, a signature of the photoexcited electron and hole populations is present at approximately 100 eV and 99 eV, respectively, as shown in the bottom panel of Figure 2A. When the Si alone is photoexcited (dashed line in Figure 2A), an increased absorption is measured below 99 eV (holes) and above 101 eV (structural). A decreased absorption is measured around 100 eV (electrons). Following photoexcitation of the Si in the junction, an increased absorption is not measured below 99 eV, while a decreased absorption is still measured at 100 eV. In other words, the spectral signature of photoexcited holes is not observed within the signal to noise of the experiment in the junction but the electron signature is present. Above 101 eV, a slightly different structural change is also measured, as would be expected when comparing the stress and strain dynamics of the free standing 200 nm Si film versus the junction. Altogether, the Si edge measurement suggests that holes have left the Si but that the electrons remain, as expected for the operation of this MOS junction (Figure 1A).

Unlike the Si edge, the metallic Ni $M_{2,3}$ edge does not closely correspond to the underlying density of states (Supplementary Figure 3B). The core-hole perturbs the final state wave function, and the resultant many-body interaction exponentially increases the number of states at energies near the Fermi level. This leads to the sharp absorption feature around 66 eV in the ground state XUV spectrum. Any photoexcited changes in the edge relate to changes in occupation near the Fermi level. Following photoexcitation of the Ni alone with 800 nm light, the $M_{2,3}$ edge shifts to lower absolute energies and increases in absorption (dashed line Figure 2B). A similar, but slightly broader, change occurs when the Ni is photoexcited in the junction. Photoexcitation of Ni alone promotes electrons from the Fermi level to a higher lying conduction band, shifting the quasi-fermi level to lower energy. The presence of more holes near the Fermi level also allows for more XUV transitions to be possible near the Fermi level. Correspondingly, a negative energy shift and gain in absorption is measured in Figure 2B. The photoexcited electrons are too high in energy to affect the many-body state near the Fermi level. In the junction, the same negative shift is measured when 800 nm light excites the Si, indicating that holes have been added to the Ni near the Fermi level. Thus while holes are not directly observed in the Si, their rapid movement into the Ni is observed. The appearance of holes on the Ni corresponds to the absence of holes measured on Si. For reference, the hole spectral signature is also observed when comparing the static ground state absorption of $Ni^0$ to $Ni^{2+}$ in NiO (Supplementary Figure 4B).

The measurements of Figure 2A and 2B indicate that holes are transferred from the Si to the Ni within the first 100 fs of optical excitation. No transient signal is measured for $TiO_2$ in this same time period, as can be seen in the colormap in Figure 1C. Instead, the dynamics in $TiO_2$ start only after 1 ps and represent a shift and broadening (Figure 2C). The Ti $M_{2,3}$ core-hole has a strong interaction with the localized Ti 3d orbitals in $TiO_2$. Little screening of the core hole occurs, and as a result, the measured spectrum is distorted from the ground state density of states by atomic multiplet splitting (Supplementary Figure 3C). When the $TiO_2$ is photoexcited alone at 266 nm, a small negative shift in the edge energy is observed. The negative energy shift occurs immediately after photoexcitation because an electron is promoted from the O 2p orbital to the Ti 3d orbital, lowering the Ti oxidation state. The added electron helps screen the strong core hole interactions on the Ti such that, even with any state-blocking effects on the transition, the overall edge shift is negative. When the junction is photoexcited, a small positive shift to larger absolute energy is observed after 1 ps. This opposite shift when the Si is photoexcited is interpreted to mean that the Ti oxidation state has increased, or that holes have been added instead of electrons. The positive and negative shifts can again be confirmed by comparing the ground state absorption of different Ti charge states (Supplementary Figure 4C). The spectral shifts assigned to holes have opposite signs in the Ni and $TiO_2$ because the predominant core-hole effects are different. The excess holes increase the Ti oxidation state, decreasing the core-hole screening and shifting the peak positively. The excess holes on the Ni perturb the many-body state at the transition edge, shifting the quasi-Fermi level and opening up new transitions.

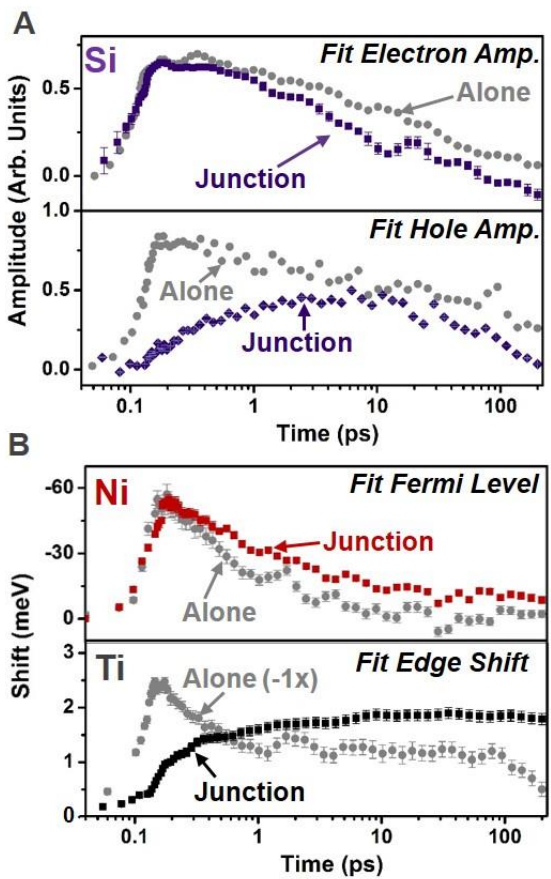

**Figure 3. Comparing the fit kinetics separately and in a junction.** The spectral signatures described by Figure 2 are fit to extract the excited state kinetics, shown here in Figure 3. In each case the error bars correspond to the non-linear-fit standard error from a robust-fit weighted by the experimental uncertainty. For each plot, the junction fit parameters are shown as the colored symbols, while the grey symbols are for the isolated material. The time scale is logarithmic with a 100 fs offset for visualization. **A.** The amplitude of electron spectral signature in the junction (purple symbols) has a similar rise time compared to the isolated material, but an increased decay rate. In the isolated Si, the amplitude of the hole spectral feature (light-purple symbols in bottom panel) follows the same kinetics as the electrons. In the junction, the hole signature slowly grows until 10 ps, suggesting that the initial photoexcited hole population was transferred out from the Si. **B. (top)** The fit Fermi level of Ni decreases when the Si in the junction is excited with 800 nm light or when a 20x thicker Ni film alone is excited with 800 nm light. The fit Fermi level then decays on a longer time scale in the junction. The amplitudes are scaled for comparison of the rise times. **(bottom)** When $TiO_2$ alone is excited with 266 nm light, a decrease in the fitted edge energy is observed because of the ligand-to-metal charge transfer. In the junction, the Ti edge fit energy increases on a time scale that matches the decay of the Fermi level in Ni.

The fit kinetics for the electrons and holes on Si, as well as the Ti and Ni edge shifts, are shown in Figure 3. The fit procedure qualitatively mirrors the discussion of Figure 2 and is described in more depth in the SI. In brief, the ground state absorption is modeled (Supplementary Figure 3) using the Bethe-Salpeter equation with density functional theory for the Si and $TiO_2$ while using an analytic many-body theory expression for the Ni. The photoexcited data are then fit at each time point as equilibrium changes to the ground state. For the Si, the electron and hole signatures are fit based on previous analysis (Figure 3A)[22,31]. For the Ni and Ti, the edge shifts are fit since no distinct spectral signature exists for the photoexcited electrons and holes (Figure 3B). For reference, all fit quantities are shown in Supplementary Figure 5. In Figure 3, the error bars of the fit process are shown on the symbols in the graphs, and the data for the material alone is shown as the grey symbols.

The fit kinetics confirm and quantify the qualitative observations of Figure 2. Specifically, following photoexcitation, holes are absent on the Si in the junction while the photoexcited electrons exist in the same magnitude as Si alone. An initial change is not measured in the Ti edge. However, an edge shift that correlates with increased holes in the Ni is measured on a sub-100 femtosecond time scale, in agreement with the expected photoexcited tunneling of holes expected for the p-type MOS junction. After a few hundred femtoseconds, the fit edge shift of the Ni decreases in magnitude while the positive edge shift of the Ti begins to rise on a similar timescale. This observation correlates with the transferred holes leaving the Ni by back-diffusing through the $TiO_2$. The fit kinetics at >100 ps are further consistent with the arrival of holes at the $Si/TiO_2$ interface, and the following injection into the Si bulk or surface recombination with the excited electrons left on the Si. This is evidenced by the similar timescales of the decay of the fit Si hole and electron amplitudes as the decay of the Ti fit edge shift.

**Discussion**

The hole transfer, diffusion, and surface recombination can be quantified using the fit kinetics of Figure 3 by taking advantage of the fact that the relative timing between the Ti and Ni or Si peaks is maintained in the spectra and fit kinetics. This is possible because at least two different elemental edges are measured simultaneously for each transient experiment. The Ti and Ni edges are measured simultaneously using Ar high harmonics and the Si and Ni edges are measured simultaneously using He high harmonics. For all fitted values, the error bars are in terms of the mean standard error as calculated through the Jacobian and covariance of a multi-start fit (MATLAB 2018b, MultiStart) of the experimental data within a 500% range of the final value. The 2 mOD experimental error is reflected by the scatter of the points on each plot.

The calculation of the charge carrier dynamics is developed as follows. First, the transit time for holes through the $TiO_2$ layer is quantified by fitting the magnitude of the Ni edge rise time to an error function that is convoluted with the 50 fs pump pulse width (Figure 4A). Convolving the fit with the pulse duration accounts for the instrument response time. This fit for the Ni-$TiO_2$-Si junction sample is compared to a separate Ni-Si sample. The Ni-Si junction is used instead of Ni alone as the reference so that any delay from carrier transport within the Si is included. The full differential absorption following photoexcitation of the Si side is shown in Supplementary Figure 6. In the Ni-$TiO_2$-Si junction, the transfer time determined by convoluting the 50 fs pulse with an error function is delayed by $50\pm2$ fs relative to the excitation pulse while in the Ni-Si junction it is delayed by $17\pm1$ fs. These error bars are smaller than the pulse width because they are the standard error of the fit, not the experimental error of the fitted rise magnitudes. The ~20 nm $TiO_2$ therefore delays the arrival of holes on the Ni by $33\pm2$ fs, which by using the measured thickness of the $TiO_2$, gives an average hole velocity of $5.8\pm0.4x10^7$ cm/s for the tunneling process. The calculated $1.5x10^5$ V/cm built-in field after photoexcitation would therefore imply a hole mobility during tunneling of 380±40 cm$^2$/V•s, similar to the accepted value for $10^{15}$/cm$^2$ p-doped Si of 450-500 cm$^2$/V•s [32].

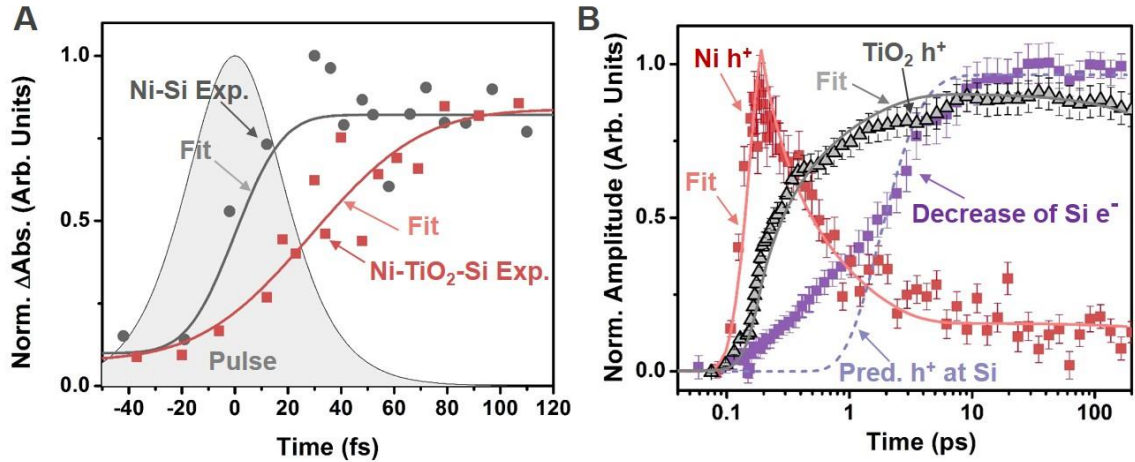

**Figure 4. Quantifying the photo-initiated hole tunneling and diffusion in the junction. A.** The square light red symbols represent the magnitude of the measured rise time of the Ni edge in the junction; their scatter represents the error of the experimental measurement. The grey circles represent the rise time of the Ni in a Si-Ni junction with no TiO$_2$ spacer. Fitting the experimental data to an error function (solid lines) convoluted with the excitation pulse gives a delayed rise of 17±1 fs for the Si-Ni and 50±2 fs with the TiO$_2$ layer. The transit time in the TiO$_2$ is therefore obtained as 33±2 fs. **B.** The edge shift kinetics, which as noted previously indicate the hole kinetics, measured for the TiO$_2$ (grey triangles) and the Ni (red squares) in the junction are compared to the increase in electron recombination (or decrease in electron signature at the Si edge) in the junction (purple squares). The solid line is a fit to the diffusion equation with a diffusion constant of 1.2±0.1 cm$^2$/s and a surface recombination velocity of 200±50 cm/s. The decrease in electron density qualitatively tracks the diffusion of holes through the TiO$_2$. The dashed line represents the predicted (Pred.) arrival of holes at the Si-TiO$_2$ interface based on the fit diffusion kinetics.

From the unchanged hole mobility in the TiO$_2$ with respect to the Si at early times, which implies no scattering of the holes in the TiO$_2$, and the lack of a measurable hole signal in the TiO$_2$ during tunneling, the hole transport through the TiO$_2$ is therefore suggested to be ballistic. The injection efficiency of the tunnel junction is also quantifiable as 42±6% from the excitation density and the fit Fermi level change in the Ni, which was itself calibrated by the Fermi level shift of Ni alone for a given photoexcitation density. This injection efficiency is reasonable for the junction[33]. It should be noted that the peak laser field intensity is 170 GW/cm$^2$, which would correlate with an electric field of $5.8 \times 10^6$ V/cm. This pulse energy is more than one order of magnitude less than that needed for optical field induced tunnel ionization or for optical field induced changes in the band structure[34–36]. Although not comparable to AC modulations at the optical field frequency, the DC dielectric breakdown values for Si and $TiO_2$ are $>10^7$ V/cm[37].

To test the accuracy of the element-specific kinetics, the hole back-diffusion rate through the $TiO_2$ layer is quantified by fitting the Ni and Ti edge shifts, which represent the hole kinetics, to the diffusion equation (Figure 4B and SI Section 3D). The fit uses the Si-to-Ni hole transfer kinetics in Figure 3A as the source and is discretized over the junction. The boundary condition is fit to a constant representing the surface hole arrival velocity, which can include both surface recombination and hole injection into the Si bulk. The fit gives a diffusion constant of $1.2 \pm 0.1$ $cm^2/s$, close to the 0.4 $cm^2/s$ value estimated in an annealed thin film of amorphous $TiO_2$ nanoparticles[38]. The surface recombination or injection velocity from the fit is $200 \pm 50$ cm/s, which is also similar to previous measurements of $>200$ cm/s[39,40]. The fit surface recombination velocity should only be taken as qualitative since the scan time range of 200 ps is too short for significant recombination to occur. This parameter has a much higher variance than the diffusion constant because the data were taken with logarithmic time points, so there are very few data points at long times. Accordingly, there is little dependence between the two fit parameters because they have such differing timescales. In Figure 4B, the measured increase in recombination for electrons on Si (purple squares) is also compared to the number of holes at the Si-$TiO_2$ interface as predicted by the fit diffusion equation (dashed purple line). Again, although a qualitative comparison, the arrival of holes at the interface predicted by the diffusion rate appears to correlate with the recombination of electrons on the Si. The primary source of error in these quantities is approximating the kinetics by a simple diffusion equation, rather than the experimental error or the fit standard error. This model is presented to give context to the measured dynamics in terms of the known values and is not intended to be absolute.

In conclusion, element-specific transient XUV spectroscopy is used to quantify the photo-initiated charge transfer in a Ni-TiO$_2$-Si junction. An initial ballistic hole tunneling from the Si to Ni is observed on a <100 fs timescale. The injection efficiency of photoexcited carriers was measured to be 42±6%. On a picosecond timescale, a back-diffusion of holes from the Ni to the Si through the TiO$_2$ is measured. As the holes arrive at the Si-TiO$_2$ interface, a decrease in the electron population is measured on the Si in the junction relative to the Si alone. Combined, these experiments confirm that the defect-rich TiO$_2$ efficiently mediates hole diffusion. The drift and diffusion values quantified from the transient XUV measurement also match previously reported values, confirming transient XUV as a highly versatile analysis tool for measuring charge transfer in multiple-element materials and junctions.

**Supporting Information**

Detailed descriptions of the sample preparation and characterization, XUV transient absorption spectroscopy via high harmonic generation, excited state XUV modeling of all three edges, and the diffusion equation model, along with Supplementary Figures 1-6, can be found in the Supporting Information.

**Acknowledgements**

The authors gratefully acknowledge financial support provided by the U.S. Air Force Office of Scientific Research (Grant No. FA9550-14-1-0154). The transient absorption measurements were done using a previously built instrument that is funded and has personnel supported (I. J. P.) by the U.S. Department of Energy, Office of Science, Office of Basic Energy Sciences, Materials Sciences and Engineering Division, under Contract No. DEAC02-05-CH11231, within the Physical Chemistry of Inorganic Nanostructures Program (KC3103). S.K.C. acknowledges support by the Department of Energy, Office of Energy Efficiency and Renewable Energy (EERE) Postdoctoral Research Award under the EERE Solar Energy Technologies Office.

**Author Contributions**

S.K.C. and S.R.L. designed the study. S.K.C., I.J.P. and A.L. performed the transient XUV measurements, and S.K.C. performed the excited state modelling and data analysis. B.R.L., B.M.M. and M.E.V. were responsible for sample fabrication, and S.S. performed the sample characterization. S.K.C., I.J.P., B.R.L., A.L., B.M.M., S.S., M.E.V. and S.R.L. wrote and revised the manuscript.

**Competing Financial Interests**

The authors declare no competing financial interests.

**Data Availability**

The data generated and analyzed during the current study are available from the corresponding author on reasonable request.

# Layer-Resolved Ultrafast XUV Measurement of Hole Transport in a Ni-TiO$_2$-Si Photoanode

Scott K. Cushing[1], Ilana J. Porter[2,3], Bethany R. Lamoureux[2], Angela Lee[2], Brett M. Marsh[2],

Szilard Szoke[1], Mihai E. Vaida[2,5], Stephen R. Leone[2,3,4*]

[1]Division of Chemistry and Chemical Engineering, California Institute of Technology, Pasadena, CA 91125, USA

[2]Department of Chemistry, University of California, Berkeley, CA 94720, USA.

[3]Chemical Sciences Division, Lawrence Berkeley National Laboratory, Berkeley, CA 94720, USA.

[4]Department of Physics, University of California, Berkeley, CA 94720, USA.

[5]Department of Physics, University of Central Florida, FL 32816, USA.

Author e-mail address: srl@berkeley.edu

## Section 1. Thin Film Growth and Characterization

The physical vapor deposition method was employed to grow $TiO_2$ and Ni films onto a 200 nm thick, <100> silicon membrane (Norcada), using home-built evaporators containing Ti (99.98 % pure, from Kurt Lesker) and Ni (99.98 % pure, from Kurt Lesker) filaments. To make the layered sample studied here, first Ti was evaporated on the silicon membrane under an oxygen (99.998% pure) atmosphere of $7.4 - 8.0$ x $10^{-9}$ Torr with the Ti filament at ~1157 °C. After 7.5 hours of Ti deposition, the oxygen was pumped off and Ni was evaporated at a temperature of ~1000 °C for 3.5 hr. A cross-sectional TEM of the sample can be seen in the main text Figure 1B. The samples with the $TiO_2$ and Ni alone were deposited on 50 nm thick diamond membranes.

## Section 2. High Harmonic Generation

The NIR pump pulses used in this experiment are a portion of the 3.5 mJ, 40 fs pulses centered at 800 nm produced by a 1 kHz Ti:Sapphire chirped pulse amplifier (Spitfire Pro, Spectra Physics). The 266 nm pump pulses used to photoexcited the $TiO2$ are produced via third harmonic generation of the 800 nm pulses. XUV probe pulses are produced by high harmonic generation of 2.5 mJ of the 800 nm pulses, a portion of which is converted to 400 nm using an in-line second harmonic generation scheme[1].This allows for the production of both even and odd harmonics. High harmonic generation occurs in a semi-infinite gas cell (40 cm) filled with either 250 Torr (approximately 3.3 x $10^4$ Pascal) helium gas for the Ni $M_{2,3}$ and Si $L_{2,3}$ edges, or 40 Torr (approximately 5.3 x $10^3$ Pascal) argon for the Ti $M_{2,3}$ edge. The residual NIR and visible light is blocked by a 0.5 mm thick, 5 µm pore size glass capillary array, which transmits the XUV onto the sample[2]. The XUV pulses transmitted through the sample are spectrally dispersed by a variable line spacing grating (35 eV-110 eV) and captured by a charge-coupled device camera (PIXIS-400, Princeton Instruments).

The XUV probe spot size is approximately 200 µm at the sample, the samples are raster scanned in 100 µm steps between each time delay, and a stream of dry nitrogen is flowed over the sample in order to dissipate heat to avoid thermal damage. Approximately 500 pulses are coadded together to produce a camera image of XUV light versus photon energy. Pump-on and pump-off camera images comprise a single time delay, with the delays spaced logarithmically after time zero (61 delay times, -2500 fs to +200 ps about time zero). Approximately 250 such scans are averaged together to produce each transient absorption measurement. Static absorbances are reported as the logarithm of the ratio of XUV photon flux between no sample and the sample. Differential absorbance is reported as the difference between the absorbance measured with the pump on versus the pump off, with outlier measurements removed with Weiner filtering. The pump spot size is approximately 250 µm at the sample, which encompasses the entire probe spot. Pump power densities are approximately 2 mJ cm$^{-2}$ for all excitations of the Si edge (junction sample, Si-Ni sample, and Si alone sample), approximately 1 mJ cm$^{-2}$ for the $TiO_2$ alone sample, and approximately 0.3 mJ cm$^{-2}$ for the Ni edge excitations (Ni alone sample and thin sample in Supplementary Figure 2B).

**Section 3. Theoretical and Computational Details.**

**A. Silicon L$_{2,3}$ Edge**

The ground state Si XUV absorption was predicted using the OCEAN code (Obtaining Core-level Excitations using Ab initio methods and the NIST BSE solver). The energy-dependent broadening is included using a Drude-Lindhard single-plasmon pole model for the electron loss function. The ground state electron densities and wave-functions are calculated at the density functional level (DFT) using Quantum-ESPRESSO[3]. The local density approximation (LDA) using a norm-conserving generalized gradient approximation (GGA) Perdew-Burke-Ernzerhof pseudopotential is used to calculate the density of states with a converged k-point mesh of 20x20x20 points and a plane wave cutoff of 100 Ry. The lattice constant is converged at 5.46 Angstroms. Projector augmented wave (PAW) reconstructed wave functions are used for calculating the core-level transition matrix elements. A real-space random phase approximation is used to estimate the dielectric screening inside a sphere around the atom, while the Levine-Louie dielectric function is used outside this sphere[4,5]. The Bethe-Salpeter (BSE) equation is then used to calculate the final electron-hole states.

In the BSE-DFT calculation, the final states are converged at k-point meshes of 8x8x8 and using a total number of bands of 100. The projector augmented wave states are converged at k-point meshes of 2x2x2 and using a total number of bands of 200. The SCF mixing is taken as 0.7 with 250 iterations used. The BSE mesh is 6x6x6, with a cut-off radius of 8.0 Bohr. The projector augmented wave shell radius is taken as 8.0 Bohr with a 0.8 scaling factor of the slater G parameter. The dielectric constant of silicon is taken as 11.7. XUV dipole orientations along the [100] and [110] directions are calculated, but within the experimental broadening, little difference is found in the final predicted x-ray absorption. A comparison of this fit with the ground state absorbance is shown in Supplementary Figure 3A.

The excited state changes to the Si L$_{2,3}$ edge are known to originate in a variety of electronic and structural dynamics. A simplified version of the model to extract electron energies, hole energies, and the temperature of the lattice is used[6]. Specifically, the differential absorption features above 101.5 eV are mainly from structural distortions related to heating the lattice, similar to what is measured in EXAFS. A multivariate regression is used to extract the amplitude of the initial and final thermal contributions to the differential absorption. The amplitude of the structural dynamics above 101.5 eV is multiplied with the full spectrum and subtracted from each time point. The residual, filtered by the known energy ranges of the electron and hole contributions, represents the electron and hole energy as a function of time. The relative amplitudes are calculated by integrating over these ranges, giving the values shown in the main text Figure 3A. All of the fitted parameters, not just the electron and hole populations shown in the main text, are included in Supplementary Figures 5A and 5B for the Si in the junction and alone, respectively. Differential absorptions for these two samples can be found in the main text Figure 1C for the Si in the junction and Supplementary Figure 1A for the Si alone.

**B. Nickel M$_{2,3}$ Edge**

The Ni M$_{2,3}$ edge is modeled using the many-body approach of Ohtaka and Tanabe[7]. This model is derived by summing over the transition probabilities for all possible final states, assuming parabolic bands and a scattering potential caused by the core hole. The resultant ground state absorption includes the "orthogonality catastrophe" or "white line" effect common to the x-ray absorption of metals,

$$I'(\omega^*) = \frac{1}{D}\left[\frac{D}{2\pi T}\right]^{\zeta_0} Re\left[e^{i(\frac{\pi}{2})(\zeta_0-1)}\frac{\Gamma\left(\frac{1-\zeta_0}{2}-\frac{i\omega^*}{2\pi T}\right)\Gamma(\zeta_0)}{\Gamma\left(\frac{1+\zeta_0}{2}-\frac{i\omega^*}{2\pi T}\right)}\right] \qquad (1)$$

where $I'$ is intensity, $\omega^* = (\omega - \omega_{th})$, which is the frequency with respect to the edge onset frequency $\omega_{th}$ as given by the Fermi level of the metal, $D$ is the bandwidth or energy difference between the Fermi level and conduction band upper edge, $T$ is temperature, and $\zeta_0$ is the phase shift of the Fermi sea caused by scattering with the core-hole contact potential. The ground state Ni $M_{2,3}$ edge is modeled by fitting the values of $\omega_{th}, T, D$ and $\zeta_0$ to the ground state absorption spectrum. To account for the spin-orbit splitting of the Ni 3p level, the total spectrum is modeled as the sum of two of such peaks, keeping all four fit variables constant, and fitting the spin-orbit splitting energy and peak ratio. A comparison of this fit with the measured spectrum is shown in Figure S3. Since the phase factor $\zeta_0$ depends on the energetics of the electrons filling the Fermi sea, it contains no new information other than $\omega_{th}$ and $T$.

To model the differential absorption spectrum of the excited state Ni $M_{2,3}$ edge, the modeled ground state spectrum for both spin-orbit components is subtracted from the modeled excited state fit, in which all variables are held constant from the ground state calculation except $\omega_{th}$, $T$, and $\zeta_0$. Since Ni is not a free-electron metal, the magnitudes of $D$, $T$ and $\zeta_0$ in the ground state are nonphysical. The magnitude of the excited state changes, however, prove experimentally accurate. For example, the temperature shift matches the predicted value for the given carrier concentration and heat capacity. The change in Fermi level matches the number of carriers photoexcited in the Ni. The differential absorption spectrum of the photoexcited Ni in the junction can be seen in the main text Figure 1C, and the differential absorption of the Ni alone is shown in Supplementary Figure 1B. A comparison of the ground state absorbances of these two samples is found in Supplementary Figure 2A. All fitted parameters are shown in Supplementary Figures 5C and 5D for the Ni in the junction and alone, respectively. A comparison of the fitted Fermi level for the junction and alone sample are in the top panel of the main text Figure 3B.

## C. Titanium $M_{2,3}$ Edge

Similar to the Si $L_{2,3}$ edge, the ground state absorption of the TiO₂ at the Ti $M_{2,3}$ edge is first modelled using the OCEAN code. The DFT k-point mesh was 6x6x4 with a lattice constant of a=3.75 and c=9.38 Angstroms. The local density approximation (LDA) is used with a norm-conserving Perdew-Wang pseudopotential since OCEAN cannot use LDA+U or hybrid functionals. A plane wave cut-off of 100 Ry was used for the pseudopotential. In the BSE calculation, the final states are converged at k-point meshes of 6x6x4 and using a total number of bands of 50. The projector augmented wave states are converged at k-point meshes of 2x2x2 and using a total number of bands of 100. The BSE mesh is 4x4x4, with a cut-off radius of 4.0 Bohr. The projector augmented wave shell radius is taken as 4.0 Bohr with a 0.8 scaling factor of the slater G parameter. The dielectric constant of TiO₂ is taken as 20 to approximate amorphous TiO₂. XUV dipole alignments along the [100] and [111] directions are calculated, but within the experimental broadening and amorphous TiO₂ layer, little difference is found in the final predicted x-ray absorption. An energy-dependent broadening was included by using a separate

Lorentzian for each atomic multiplet split peak. The value of the broadening was 3 eV except for the central peak, which was 5 eV. This modeled spectrum is compared to the measured static absorbance in Supplementary Figure 3C.

The atomic multiplet splitting in the metal oxide means that only charge state and a change in broadening can be extracted as a function of time. This is achieved by fitting a global Lorentzian broadening and an energetic shift to the ground state absorption, and then subtracting the unmodified ground state absorption to calculate the differential absorption. These fitted parameters are shown in Supplementary Figure 5E for the TiO₂ in the junction and in Supplementary Figure 5F for the TiO₂ alone. Differential absorption following photoexcitation for the TiO2 in the junction and alone can be found in the main text Figure 1C and Supplementary Figure 1C, respectively. A comparison of the fitted edge shift for both samples is shown in the bottom panel of the main text Figure 3B.

## D. Fitting Routines

The fitting routines reported in the discussion section of the main text and shown in the main text Figure 4 all use a robust, non-linear fit procedure. Error bars reported represent the standard error of the fit, which includes cross terms between all fit parameters. Additionally, a multi-start fit procedure was performed on all fitted parameters within a range of 500% of the final value, and the resulting parameter variances were within bounds of the standard error, as expected. The multi-start fit was performed using the MATLAB MultiStart algorithm. For the rise time fit, an error function is used, fitting the amplitude and time constant to the experimental data, as shown in the main text Figure 4A. For the diffusion equation fit, $\frac{\partial n_h}{\partial t} = \frac{\partial}{\partial z}\left(D_h \frac{\partial n_h}{\partial z}\right) + S(t,z)$, the one dimensional diffusion equation is solved with the diffusion constant ($D_h$) and surface recombination/injection velocity at the Si-TiO₂ interface held as fit parameters, as shown in the main text Figure 4B. The hole population is represented by $n_h$. The source of holes $S(t,z)$ is the error function fit to the Ni rise time and taken as one boundary condition. The other boundary condition includes the surface recombination/injection velocity to replicate the interface with Si. Since the hole diffusion and injection/recombination occur on such differing timescales, these parameters have very little crosstalk.

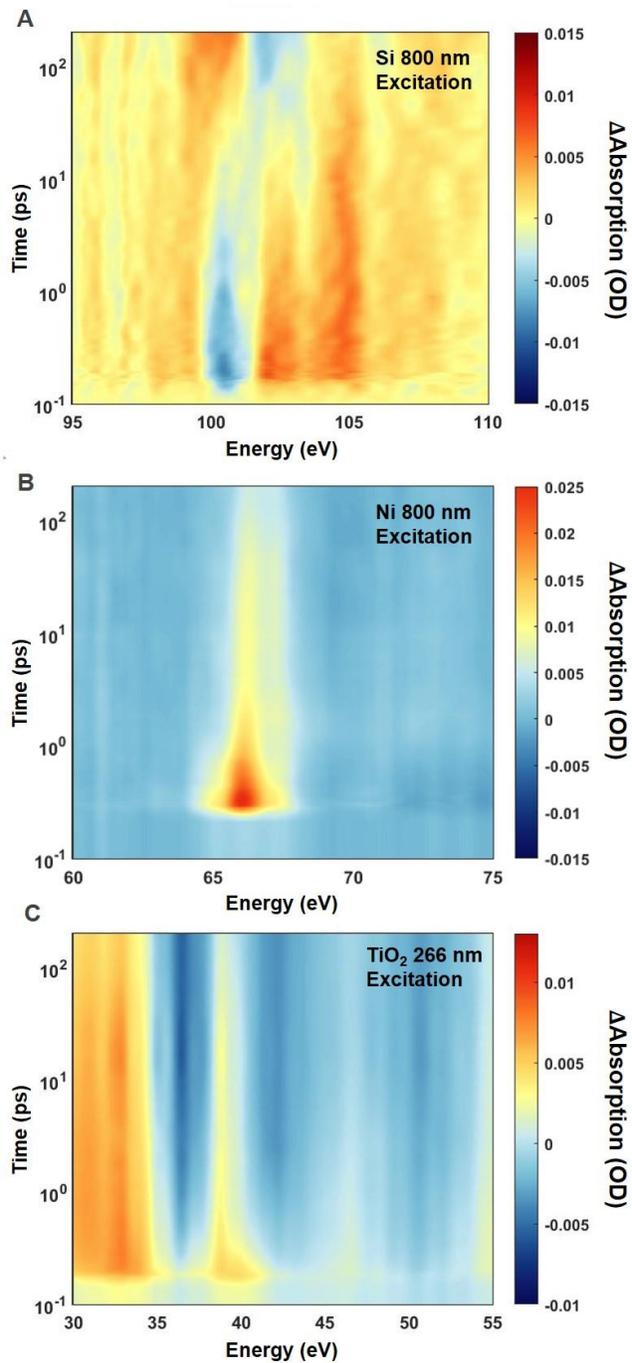

**Supplementary Figure 1. Transient Differential Data for Reference Samples.** The differential XUV absorption is shown on a negative (blue) to positive (red) colormap. The timescale is logarithmic and offset by 100 fs for visualization. **A.** Differential absorption for Si alone with 800 nm light. **B.** Differential absorption for photoexcitation of Ni alone with 800 nm light. **C.** Differential absorption for 266 nm photoexcitation of $TiO_2$ alone.

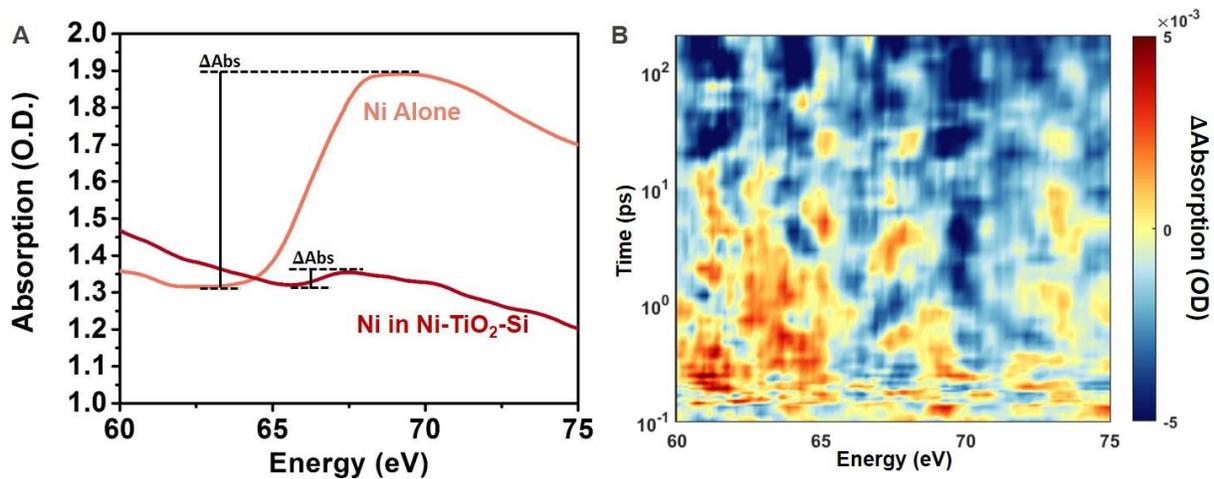

**Supplementary Figure 2. Ni in Ni-TiO₂-Si versus Ni Alone. A.** The absorption of the Ni in the junction is over ten times less than the Ni alone film when measured from pre-edge dip to peak ($\Delta Abs$). The lines have been vertically offset for comparison. **B.** The differential absorption following 800 nm excitation of a thin Ni sample with similar thickness to the Ni in the junction. This thin Ni film has no signal within a <5 mOD signal to noise ratio. The thicker Ni alone sample has an ~25 mOD signal as seen in Figure S1B.

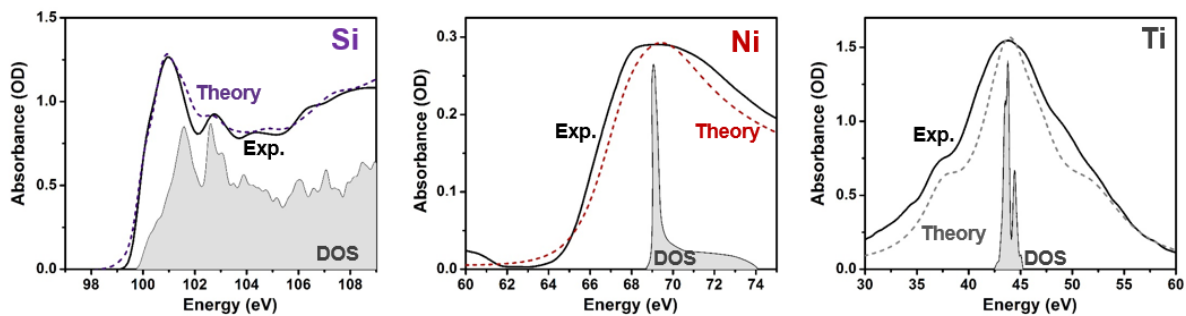

**Supplementary Figure 3. Modeling the XUV ground and excited state absorption. A-C.** The experimental XUV absorption of each element (solid lines) is compared to the BSE-DFT calculation (dashed lines) and the valence density of states (DOS) for each material. How well the valence charge density screens the core-hole perturbation from the XUV excitation determines to what degree the XUV spectrum reflects the valence density of states.

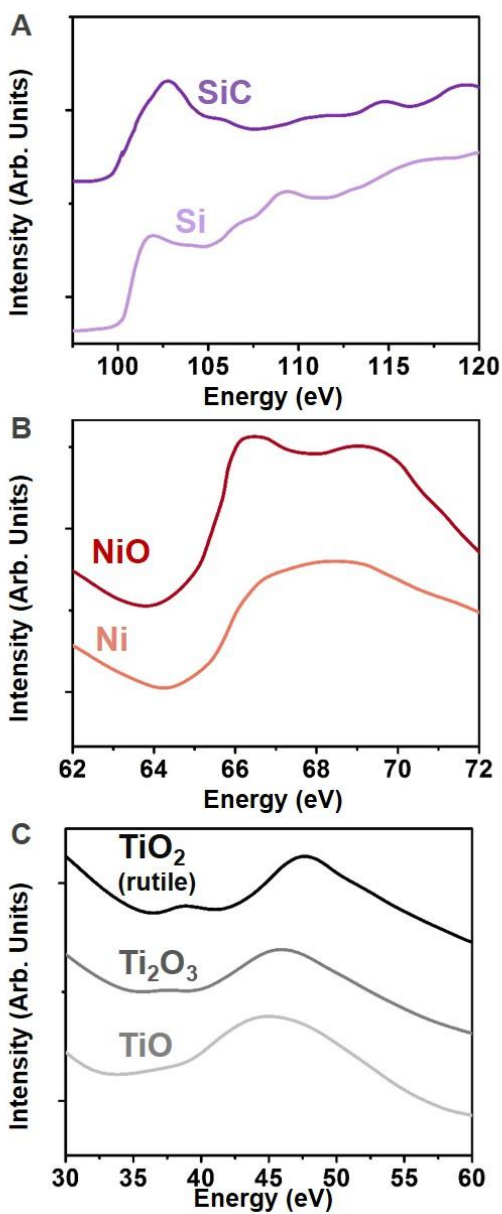

**Supplementary Figure 4. Comparison of EELS Absorption in the XUV Range for Different Compounds.** XUV absorption data from synchrotron sources is not readily available, so the electron energy loss (EELS) absorption is shown for each element[8]. The EELS absorption process involves a one-electron excitation instead of a one-photon excitation **A.** For Si, the core-hole is well screened, and the peak structure and energy are more representative of the underlying band structure than the oxidation state (compare Si and SiC). **B.** For Ni, the edge rise is sensitive to the number of holes or oxidation state. When electrons are removed in the metal, the Fermi level changes, and the edge shifts to lower energy while increasing in absorption. **C.** For Ti, the broad absorption and appearance of multiplet split peaks is sensitive to the oxidation state.

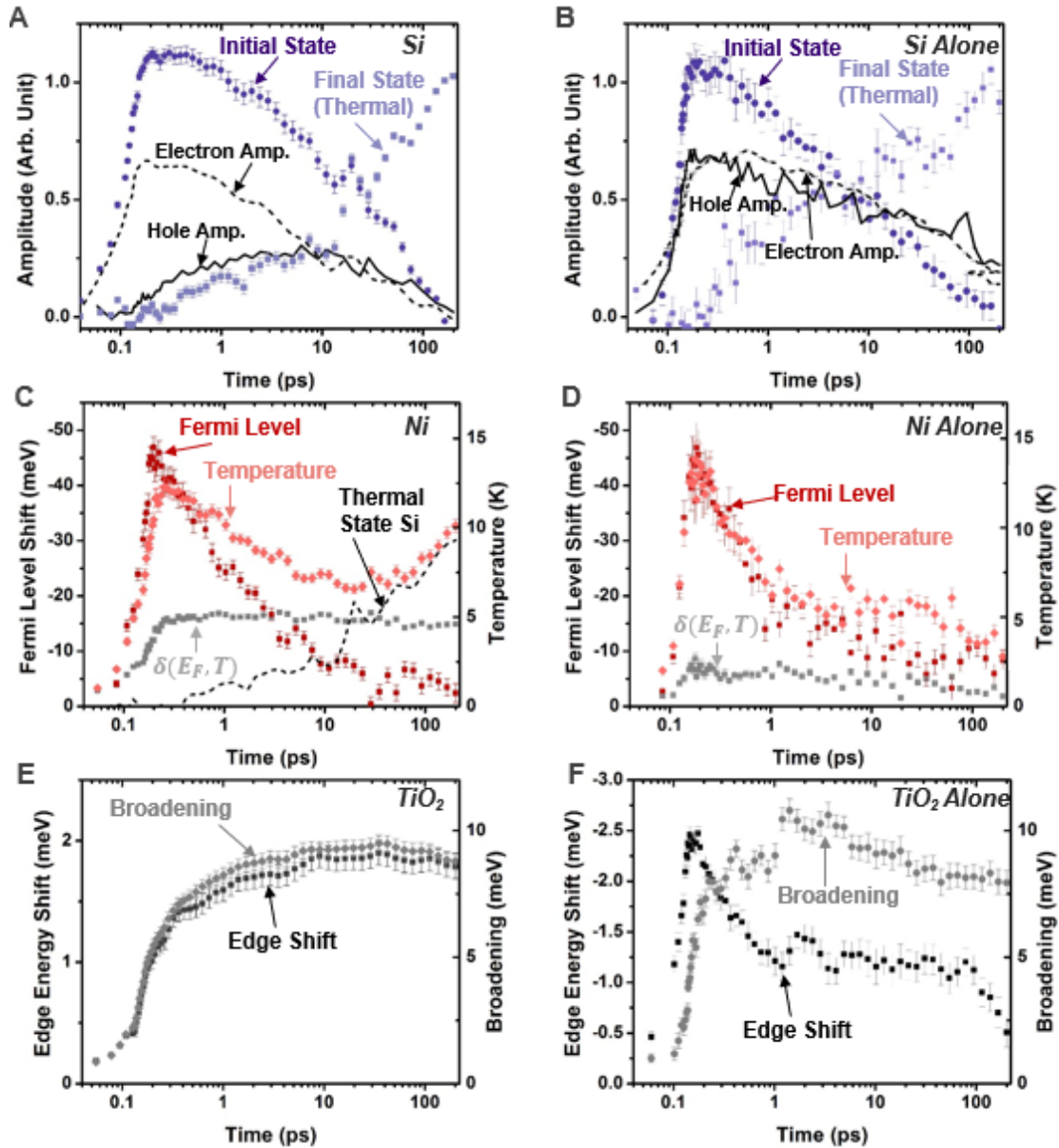

**Supplementary Figure 5. All Fit Parameters for each XUV Edge as a Function of Time. A.** For Si in the junction, the amplitude of the initial and final states used in the multivariate regression are shown as the dark purple circles and light purple squares, respectively. The final state represents a heated lattice. The electron and hole features' spectral amplitudes after removal of the final thermal state are shown as the dashed and solid black lines, respectively. **B.** The same as in Panel A, but for Si alone. **C.** For Ni in the junction, the fit Fermi level (red circles), temperature in the Fermi-Dirac distribution (light red diamonds), and the phase factor (gray squares) which is co-dependent on the Fermi level and temperature are shown. The final thermal state from the Si fit is shown as a dashed black line. **D.** The same as in panel C but for Ni alone. **E.** For $TiO_2$, the fit broadening (grey diamonds) and edge shift (black squares) of the Ti edge. The error bars correspond to the non-linear-fit standard error from a robust-fit weighted by the experimental uncertainty. **F.** The same as in Panel E but for $TiO_2$ alone.

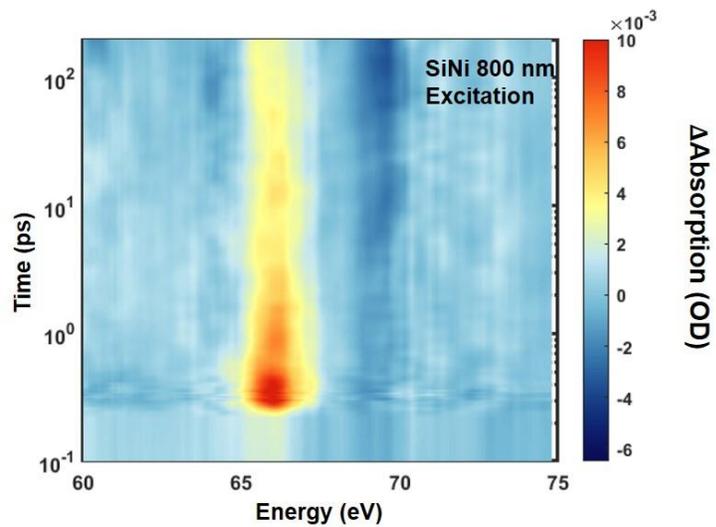

**Supplementary Figure 6. Transient Differential Data for 800 nm Photoexcitation of Si-Ni as Measured at the Ni Edge.**